\documentclass[aip,reprint]{revtex4-2}

\usepackage{nicefrac}
\usepackage{bm}
\usepackage{graphicx}
 \usepackage{amssymb}
 \usepackage[usenames,dvipsnames]{xcolor}
 \usepackage{upgreek}
\usepackage{ulem}

\DeclareTextSymbol{\degree}{T1}{6}

\newcommand{\e}{\mathrm{e}}
\newcommand{\deriv}{\mathrm{d}}

\draft 

\begin{document}




\title{Phase-contrast imaging of a dense atomic cloud}




\author{\firstname{M.} \surname{Frometa Fernandez}}
\affiliation{Instituto de F\'isica de S\~ao Carlos,
Universidade de S\~ao Paulo, S\~ao Carlos, SP 13566-970, Brazil}

\author{\firstname{P. G.} \surname{Santos Dias}}
\affiliation{Departamento de F\'isica, Universidade Federal de S\~ao Carlos, S\~ao Carlos, SP 13565-905, Brazil}

\author{\firstname{P. H.} \surname{Nantes Magnani}}
\affiliation{Departamento de F\'isica, Universidade Federal de S\~ao Carlos, S\~ao Carlos, SP 13565-905, Brazil}

\author{\firstname{M.} \surname{do Amaral Martins}}
\affiliation{Departamento de F\'isica, Universidade Federal de S\~ao Carlos, S\~ao Carlos, SP 13565-905, Brazil}

\author{\firstname{M.} \surname{Hugbart}}
\affiliation{Universit\'e C\^ote d'Azur, CNRS, INPHYNI, France}

\author{\firstname{A.} \surname{Cipris}}
\affiliation{Instituto de F\'isica de S\~ao Carlos,
Universidade de S\~ao Paulo, S\~ao Carlos, SP 13566-970, Brazil}

\author{\firstname{Ph. W.} \surname{Courteille}}
\affiliation{Instituto de F\'isica de S\~ao Carlos,
Universidade de S\~ao Paulo, S\~ao Carlos, SP 13566-970, Brazil}

\author{\firstname{R.} \surname{Celistrino Teixeira}}
\affiliation{Departamento de F\'isica, Universidade Federal de S\~ao Carlos, S\~ao Carlos, SP 13565-905, Brazil}
\email{teixeira@df.ufscar.br}


\date{\today}

\begin{abstract}

We present the experimental production and characterization of a dense cold atomic cloud of \(^{88}\text{Sr}\) atoms, optimized for the future studies of light transport in highly dense regimes. Using narrow-line molasses on the 689 nm transition, combined with a far off-resonant optical dipole trap, we achieve spatial densities as high as \(7.9 \times 10^{13} \, \text{atoms/cm}^3\) and optical depths up to 64. This approach stands out from previous methods by integrating narrow-line molasses with an optical dipole trap, enabling high-density samples without relying on evaporative cooling. Unlike traditional absorption imaging, which becomes inaccurate in such dense regimes, we demonstrate that phase-contrast imaging (PCI) can reliably reconstruct the in-situ density profile even for highly spatially and optically dense samples. The use of a spatial light modulator instead of a fixed phase plate in the PCI setup provides enhanced flexibility and control of imaging parameters, making this imaging technique robust against imaging artifacts and adaptable to varying experimental conditions.
Moreover, we derive theoretical conditions for reliable PCI operation in dense regimes and validate these experimentally,  showing excellent agreement with time-of-flight measurements even at the highest densities. Our results establish a robust method for producing and characterizing dense atomic clouds.

\end{abstract}

\pacs{}

\maketitle 

\section{Introduction}

The linear interaction of a macroscopic material with coherent light is fully encapsulated in its linear electrical susceptibility, which characterizes both the phase dispersion and the extinction coefficient of the propagating light. 
The susceptibility is a function of the light frequency, and must be obtained experimentally for all materials \cite{HandbookOptical1997}. On the other hand, at the opposite limit of single atoms or molecules interacting with light, quantum electrodynamics provides very precise theoretical predictions for the polarizability of the individual constituents of matter, extensively verified experimentally \cite{Werij1992, Yasuda2004}. For a very dilute gas, the susceptibility can be described in terms of individual polarizabilities using the Lorentz model for electrical susceptibility\cite{Jackson1999}. However, this approach fails when applied to solids, liquids, or gases at room temperature and atmospheric pressure \cite{Javanainen2016, Andreoli2021}. Currently, there is no known bottom-up approach for bridging the gap between the macroscopic susceptibility and the microscopic polarizability of the individual constituents for everyday material systems; in particular, the Lorentz model shows orders-of-magnitude discrepancies with respect to the experimentally obtained values for such systems. Those discrepancies most probably stem from strong interactions among the individual constituents, invalidating any mean-field procedure.

Dense samples of cold atoms are an ideal test bench for studying the intrinsic many-body problem of the relation between the individual polarizability and the macroscopic electrical susceptibility of the sample. Recent theoretical results\cite{Javanainen2016, Andreoli2021} point to many-body modifications of the macroscopic susceptibility for densities achievable by cold atoms experiments. Experimental evidences of those discrepancies were verified in cold Rubidium setups \cite{Chomaz2012, Jennewein2018}, but remaining difficulties in developing a theoretical model for explaining them ask for new tests with atoms of simpler atomic structure, such as bosonic Strontium atoms. Bosonic Strontium presents optical dipolar transitions with the simplest possible sublevel structure, from the fundamental level of total angular momentum $J = 0$ to an excited level of angular momentum $J = 1$, being ideal for such studies. 

The first significant contribution of this work is the development of an experimental protocol for producing dense and cold samples of $^{88}$Sr atoms to probe many-body modifications of the macroscopic susceptibility of matter. Since $^{88}$Sr atoms have a very small s-wave scattering length\cite{Stellmer2013} ($-2 a_0$, with $a_0$ the Bohr radius), we can expect that atomic interactions and three-body relaxation processes will not hinder the production of samples of high densities. However, traditional evaporative cooling techniques are ineffective for further increasing the spatial density of these samples. Instead, we show that applying an optical molasses with light quasi-resonant to the narrow $689~$nm transition of Strontium, while the atoms are confined by a far off-resonant optical dipolar trap, allows them to accumulate at the bottom of the tight optical trap, achieving high spatial density without the need for evaporative cooling. We discuss in this paper the difficulties and advantages related to the narrow width of the molasses optical transition, and show its effectiveness in producing high atomic densities.

In the path to the production of dense samples, it became essential to equip our experiment with a reliable density measurement protocol. In the field of cold and ultra-cold atom optics, imaging the atomic ensemble is a fundamental tool for probing the density and temperature of the sample, either of the whole atomic distribution, or of a specific fraction of the cloud in a different internal state through state-resolved imaging \cite{Stellmer2011}.
Imaging a cold atomic cloud typically relies either on measuring the impact of the atoms on quasi-resonant probe light passing through the cloud, or measuring the light scattered from the probe. Both observables are impacted by the many-body modifications of the electrical susceptibility of the cloud, that can be at play for densities comparable to, or higher than, 1 atom per volume of $\lambda^3$, with $\lambda$ the wavelength of the probe light. 
In this article, we show that the commonly used absorption imaging protocol in cold atom experiments is significantly hindered by such modifications, preventing reliable inference of the atomic cloud's density profile. Alternatively, the phase contrast imaging (PCI) technique \cite{Meppelink2010,Andrews1997} provides a signal that scales linearly with atomic density, even for densities well above 
$\lambda^{-3}$ provided that certain conditions—related to light detuning and cloud geometry, which we derive—are satisfied.
We also describe the implementation of a PCI setup in our experimental system, where we have replaced the commonly used passive phase plates by a Spatial Light Modulator (SLM). The use of the SLM offers distinct advantages, notably its ability to dynamically adjust the phase mask in real-time, providing a level of control and flexibility that is not achievable with traditional passive plates. This adaptability allows us to optimize the phase mask parameters according to varying experimental conditions, such as changes in atomic cloud geometry and density. By incorporating this flexibility and dynamic control, we can characterize the PCI technique based on the main characteristics of the phase plate design. We demonstrate that PCI enables reliable and precise density measurements, even in regimes of high spatial density and optical depth, where many-body effects influence the transport of resonant light.

This article is organized as follows. In Sec.~\ref{sec:exp_seq} we describe our experimental machine, in particular the different techniques that allow us to produce dense atomic samples without relying on evaporative cooling. Afterwards, in Sec.~\ref{sec:PCI_theory} we derive the conditions necessary to extract the density of the atomic cloud from the PCI. Then, in Sec.~\ref{sec:PCI_results} we characterize our PCI setup, showing that we can reliably obtain the density profile of the atoms even in the high density regime. We discuss our results in Sec.~\ref{sec:conclusion}.

\section{Production of a dense atomic cloud of Strontium atoms}
\label{sec:exp_seq}

\subsection{Initial trapping and cooling within magneto-optical traps}

\begin{figure} [h!]
\centering
\includegraphics[width=\columnwidth]{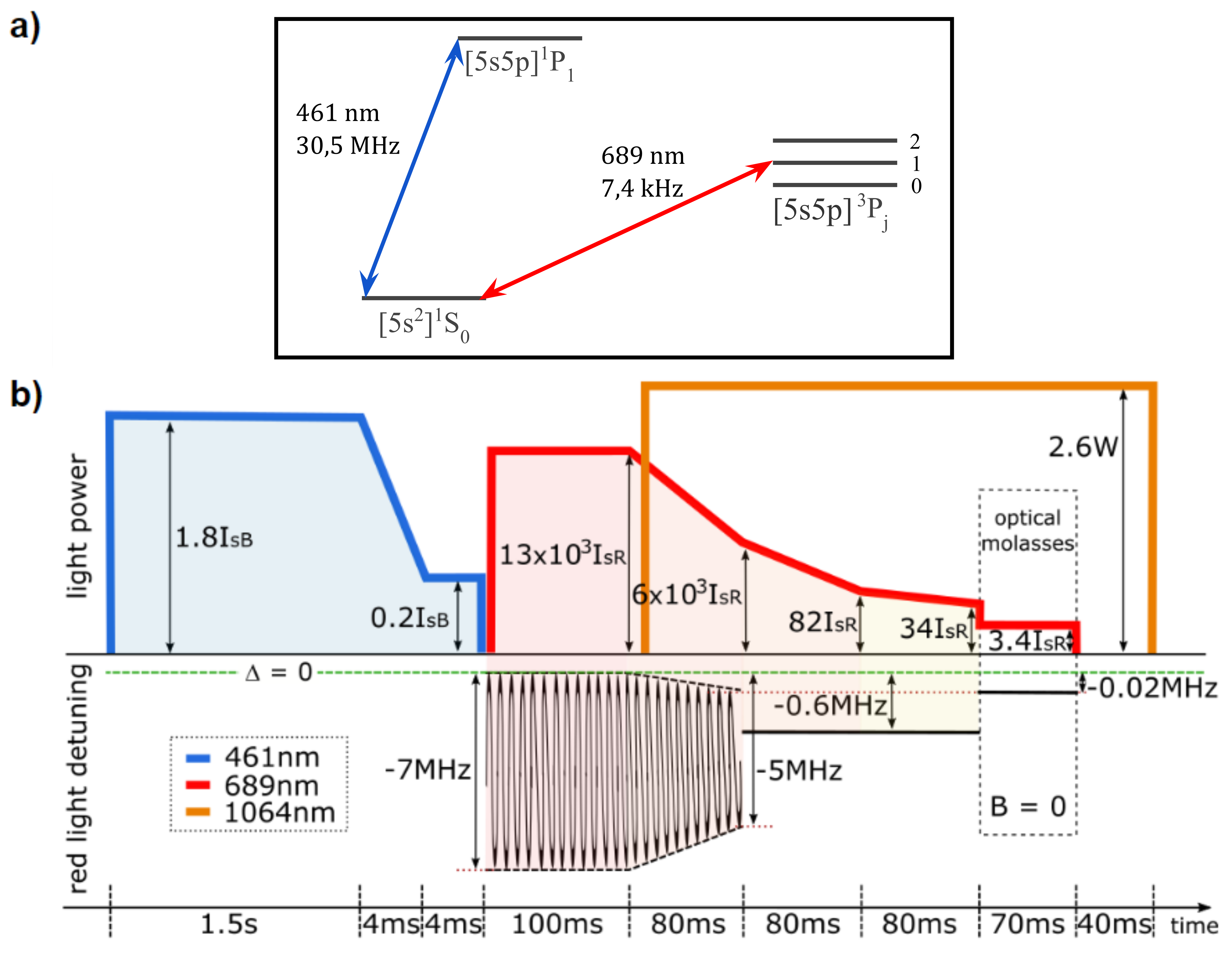}
\caption{\label{fig:experiment} (a) Sketch of the Strontium optical transitions addressed by the cooling process. (b) The experimental sequence for producing the final atomic cloud, depicting the total power of the 461 nm light used for the blue MOT (in blue), the total power of 689 nm light for the red MOT and red molasses (in red), the total 1064 nm light power for creating the optical dipole trap (ODT) (in orange), and the detuning of the 689 nm light during different cooling and trapping stages (in black). The graphics are not drawn to scale. Further details can be found in the main text.}%
\end{figure}

At the oven section of our experiment, metallic strontium is heated to $550^\circ$C and escapes through microtubes to form a collimated beam, before entering the Zeeman slower region. The Zeeman slower operates on the broad $\lambda_0 = 461~$nm Sr transition $5\text{s}^2~^1\text{S}_0 \rightarrow 5\text{s}5\text{p}~^1\text{P}_1$ of width $\Gamma = 2 \pi \times 30.5~$MHz (see also Fig.~\ref{fig:experiment}(a)), creating a flux of atoms with estimated speed $\lesssim 50~$m/s that arrives at the center of the science chamber\cite{Dias2021}. Inside the chamber, the atoms are trapped and cooled in a magneto-optical trap (MOT) whose optical force is created by three mutually orthogonal pairs of counter-propagating laser beams (of waist 5.6 mm) detuned by $\sim - \Gamma$ from the $461~$nm transition. At the end of a loading time of $1.5~$s this blue MOT contains $6 \times 10^7$ atoms. Then, as depicted in the experimental sequence in Fig. 1(b), following a fast compression of the blue MOT lasting 4 ms — during which the total light intensity is reduced from $1.8 \,\text{I}_{\text{sB}}$ to $0.2 \,\text{I}_{\text{sB}}$ (where $\text{I}_{\text{sB}} = 40.6 \ \text{mW}/\text{cm}^2$ is the saturation intensity of the blue transition) — the atomic temperature is reduced to 4 mK with minimal atom loss.
Next, the 461 nm light is turned off, and six 689 nm laser beams (of waist 5.1 mm), overlapped with the blue MOT laser beams, are turned on to create the so-called \textit{red MOT}. This MOT operates on the narrow  $\lambda_{\text{r}} = 689~$nm Sr transition $5\text{s}^2~^1\text{S}_0 \rightarrow 5\text{s}5\text{p}~^3\text{P}_1$ of width $\Gamma_{\text{r}} = 2 \pi \times 7.4~$kHz and saturation intensity $\text{I}_{\text{sR}} = 3.04 \ \upmu\text{W}/\text{cm}^2$.
To create the red MOT, we also reduce the gradient of the magnetic field along the anti-Helmholtz coils axis from 55 G/cm to 6.5 G/cm in 30 $\upmu$s, and modulate the detuning of the $689~$nm light at a frequency of 35 kHz. This modulation effectively creates a multichromatic MOT light, which increases the velocity capture range of the red MOT. This first stage of the red MOT lasts 100 ms, resulting in $4\times 10^7$ atoms with a temperature of 25 $\upmu$K. A second multichromatic stage of 80 ms reduces the amplitude of the frequency scan and compresses the cloud through a power ramp. Finally, the red MOT is operated at a single frequency, and further reduction of detuning and power with additional ramps produces a final red MOT with $1,5\times 10^7$ atoms and a temperature of 2 $\upmu$K. All details on detunings and total power used for the red MOT are given in Fig.~\ref{fig:experiment}(b).

\subsection{Far off-resonant optical dipolar trap}

During the cooling stages of the red MOT, we turn on the conservative dipolar optical trap formed by two orthogonal $1064~$nm far-detuned light beams collimated at the atoms position, to create a crossed optical dipolar trap (ODT). The horizontal ODT beam, which propagates perpendicularly to gravity (defining the vertical direction),
has a cylindrical profile with waists of $240~\upmu$m and $17~\upmu$m (the latter along the vertical direction) and a power of 1.8 W. The vertical ODT beam (propagating along the vertical direction) is circular, with a waist of $48~\upmu$m and a power of 0.75 W. The polarizations of the ODT beams are $\pi$ for the horizontal beam and $\sigma^+$ for the vertical beam, taking upward (opposite to gravity) as the quantization direction, corresponding to the direction $+z$ in the following. The trapping frequencies are $(\omega_x, \omega_y, \omega_z) = 2 \pi \times (218,\ 213,\ 694) \ \text{s}^{-1}$. The frequency $\omega_z$ is measured by the observation of atomic oscillations in the trap after a fast blink of the potential, while $\omega_x$ and $\omega_y$ are calculated from a simulation of the trapping potential based on the measured size and power of the trapping beams. Although the ODT is turned on during the red MOT cooling stages, the atoms are loaded into the ODT only at the end of the red MOT, when their positions and temperature facilitate the accumulation of the slowest atoms at the bottom of the trap.

\subsection{Narrow-line molasses combined with optical dipole trap}
The mode volume of the crossed ODT being much smaller than the final volume of the red MOT, we found that this transfer heats the cloud and limits its final density; Most of the atoms end up trapped in the shallower potential created by the horizontal ODT beam, while a faint cloud is produced at the crossing region of both ODT beams. In order to improve the transfer to the crossed ODT, we implement an optical molasses immediately after the red MOT stage, overlapped with the ODT\cite{Grimm2000optical} (see Fig.~\ref{fig:experiment}(b)), by simply turning off the magnetic field gradient while maintaining the three pairs of counterpropagating $689~$nm laser light beams of the red MOT. Without the magnetic field gradient, the counterpropagating red-detuned light produces no trapping potential, but maintains an optical molasses that continues to cool the atoms from the shallow horizontal ODT while they accumulate at the bottom of the crossed ODT.

\subsubsection{Cancellation of magnetic field}
As the molasses is produced using a narrow transition\cite{Chalony2011doppler}, its implementation requires a very well controlled magnetic field. If the Zeeman shift of the magnetic sublevels is comparable to or greater than the transition width, it can compromise the effectiveness of the optical molasses. The fundamental level of the red transition has zero total magnetic moment, while the magnetic sublevels of the excited level are shifted by $\Delta_Z = 2 \pi \times \beta_B B\, m_J$, with $\beta_B = 2.1~$MHz/G; this means that a magnetic field of $3.5~$mG already creates a Zeeman shift of $\Gamma_{\text{r}}$.

\begin{figure}
\centering
\includegraphics[width=\columnwidth]{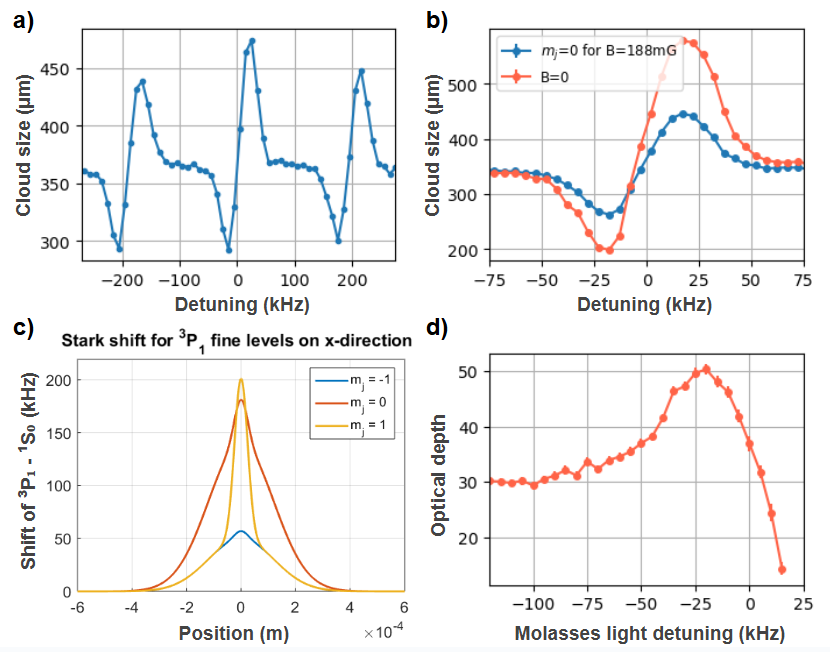}
\caption{\label{fig:molasses} (a) Horizontal radius of the atomic cloud, after 24~ms of optical molasses on the narrow 689~nm transition, as a function of the detuning of the 689~nm light, in presence of a constant magnetic field. The appearance of three dispersive-like curves reveals the effect of the molasses close to resonance to the excitation of each magnetic sublevel $m_J = \{-1, 0, 1\}$. (b) Black points and line: Same as (a), but with magnetic field cancellation. Red points and line: Without magnetic field cancellation. (c) Calculated stark shift of the different magnetic sublevels of the $5\text{s}^2~ ^1\text{S}_0 \rightarrow 5\text{s}5 \text{p}~ ^3\text{P}_1$ molasses transition. (d) Optical depth of the ODT as a function of the detuning of the red molasses.}%
\end{figure}

The cancellation of the magnetic field is done by probing its effect on the molasses itself, but with the ODT turned off to avoid any optical shift on the $689~$nm transition. Without the ODT, the atoms fall due to gravity while interacting with the molasses for $24$~ms, and the horizontal size of the cloud after this fall indicates whether the molasses heated, cooled or had no influence on the cloud (we do not consider the vertical size because the impact of gravity complicates the interpretation of the results). In fig.~\ref{fig:molasses}(a), we show the horizontal size of the atomic cloud after the molasses stage as a function of the frequency of the $689~$nm light in the presence of a constant magnetic field. The three dispersive-like curves correspond to transitions to the magnetic sublevels of the $^3\text{P}_1$ state. Each dispersive-like curve is the result of the molasses acting on the atoms, which cools (resp. heats) the cloud when tuned to the red (resp. blue) side of the resonance, decreasing (resp. increasing) its size. The resonance position is near the zero-crossing of the dispersive-like spectrum, with each resonance exhibiting a width of 40 kHz due to Doppler and power broadening. By controlling the uniform magnetic field at the atoms' position using three pairs of rectangular coils in a quasi-Helmholtz configuration placed around the science chamber, we iteratively cancel the magnetic field in all three spatial directions. This removes the splitting of the sublevels of the state $^3\text{P}_1$, and the three dispersive curves converge into a single curve at the position of the central excited level $m_J = 0$, as shown in Fig.~\ref{fig:molasses}(b). We estimate from this measurement that the magnetic field is canceled to a precision better than $10~$mG. In this situation, the optimal cooling condition occurs with a detuning of -20 kHz relative to the atomic transition resonance. The fact that the compensated molasses has the same width as the molasses acting only at the magnetically insensitive transition to the $m_J = 0$ level shows that inhomogeneities of the magnetic field are negligible.

\subsubsection{AC Stark shift}
Since the molasses is meant to operate superimposed to the ODT, we now turn our attention to the impact of the AC Stark shift caused by the ODT light on the molasses. Fig.~\ref{fig:molasses}(c) shows the AC Stark shift caused by the crossed ODT on each transition to the magnetic sublevels of the $5\text{s}5\text{p}~^3\text{P}_1$ level at $689~$nm, taking $+z$ as the quantization direction. Since the Stark shift is much bigger than the width of the molasses transition, and considering the spatial dependence of the ODT power, we see that the molasses light will be acting differently on atoms at different positions within the ODT. The molasses stage lasts for the first 70 ms of the ODT, with an intensity of $3.4\,\text{I}_{\text{sR}}$, which are the values that result in a final cloud with the highest optical depth and lowest temperature. Fig.~\ref{fig:molasses}(d) shows the optical depth of the dense cloud as a function of the detuning of the molasses light with respect to the bare atomic resonance. We see that the optimum is obtained for a detuning of $-20~$kHz, which is the same optimal value found for molasses cooling in absence of the ODT.
Further to the blue of the $689~$nm resonance, we observe some effect of the molasses, that we attribute to the AC Stark-shifted transitions to the blue, but they produce a less dense cloud.
This observation reinforces the hypothesis that the main effect of the molasses is through the cooling of the atoms as they enter the deeper, compressed potential. As it can be seen in Fig.~\ref{fig:molasses}(d), the optical depth is increased by $65~\%$ in the presence of the molasses. This result confirms the effectiveness of this approach. 

It is interesting to note that we probably have an advantage in operating the optical molasses on a narrow transition, in addition to the smaller Doppler temperature. The radiation pressure from photons that are rescattered within the cloud —typically limiting the maximum achievable densities \cite{Sesko1991,Townsend1995} — will be reduced due to the Doppler spread of the atomic resonances, since in this case not all neighbours are resonant with the scattered photons of a nearby atom. Even for a cloud temperature of $1~\upmu$K, the half-width Doppler broadening of the $689~$nm resonance is $\Delta_D = 2 \pi/\lambda_{\text{r}} \sqrt{k_B T/m} = 2 \pi \times 14~$kHz, with $k_B$ the Boltzmann constant and $m$ the mass of a $^{88}$Sr atom, which is almost twice the natural width $\Gamma_{\text{r}}$ of the atomic transition.

\subsubsection{High spatial and optical densities}

After applying the optical molasses, and optimizing the other experimental ramps, we obtain a cloud with $4.1 \times 10^5$ atoms, a resonant optical depth of $b_0 = 64$ (in the $z$ propagation direction of the imaging light), cloud radius at $1/ \sqrt{e}$ of $(R_x, R_y, R_z) = (10.2, 10.5, 3.1)~\upmu$m (corresponding to a mean transverse cloud size in the $xy$ plane of $R_r = 10.3~\upmu$m), with spatial density at the center $\rho_0 = 7.9 \cdot 10^{13}~\text{cm}^{-3}$ (in units of $1/ \lambda_0^3$: $\rho_0 \lambda_0^3 = 7.8$), and a temperature of $1.1~\upmu$K. The parameters $R_x$, $R_y$ and $b_0$ are measured from the \textit{in-situ} PCI imaging, which will be described below, while $R_z$ and the spatial density $\rho_0$ are calculated from these measured parameters, along with the temperature of the cloud obtained after time-of-flight and the measured trap frequency $\omega_z$.
\\

\section{Theoretical treatment of the imaging of a cold atomic sample}
\label{sec:PCI_theory}

In order to characterize the \textit{in-situ} density profile of our cloud, we must choose an adapted imaging protocol. In this section, we theoretically discuss how the high optical depth and spatial density of our cloud impact the different imaging strategies, and conclude that the phase contrast imaging can provide a reliable density measurement.

\subsection{The linear electrical susceptibility of a cold atomic gas}

We focus here on the simple case of an atom with a dipolar transition from a fundamental level of total angular momentum $J = 0$ to an excited level of total angular momentum $J = 1$ (for a general treatment that considers atoms with hyperfine structure, see \cite{Meppelink2010}). This applies to the transition $^1\text{S}_0 \rightarrow \nolinebreak ^1\text{P}_0$ used for imaging $^{88}\text{Sr}$ atoms in our experiment. The linear atomic polarizability of this dipolar transition (for a light intensity much lower than the saturation intensity of the transition) is given by 
\begin{equation}
    \alpha(\Delta) = i\frac{6 \pi c^3 }{\omega_0^3}\frac{1}{1-2 i\Delta/\Gamma} \ ,
\end{equation}
with $c$ the speed of light, $\omega_0$ the natural frequency of the transition, $\Delta = \omega - \omega_0$ the detuning of the laser light of frequency $\omega$ and $\Gamma$ the natural linewidth of the transition. From a macroscopical point of view, the effect of a linear, isotropic medium on the light propagation is fully characterized by the index of refraction of the medium $n(\bm{r})$ which for a dilute cloud is
\begin{equation}
    n^2(\bm{r},\Delta) = 1 + \rho(\bm{r}) \alpha(\Delta)\ ,
    \label{eq:IndexRef}
\end{equation}
where $\rho(\bm{r})$ is the atomic density of the sample. The limit of a dilute cloud corresponds to $\left|n^2 - 1 \right|\ll 1$. We define a dimensionless density $\eta(\bm{r})$ expressed as

\begin{equation}
    \eta(\bm{r}) = \rho(\bm{r}) |\alpha (\Delta = 0)|\ .
\end{equation}

Since $|\alpha|$ is maximum for $\Delta = 0$,  we consider the cloud to be dilute whenever $\eta(\bm{r}) \ll 1$. We can rewrite this condition for the peak spatial density of the sample $\rho_0 = \max_{\bm{r}} (\rho(\bm{r}))$ as
\begin{equation}
    \rho_0 \ll \frac{4 \pi^2}{3} \lambda_0^{-3} \ ,
    \label{eq:dense_regime}
\end{equation}
where $\lambda_0 = 2 \pi c/\omega_0$. This means that an atomic cloud with much less than one atom per cubic wavelength of light is dilute for any frequency of light. We note that the cloud produced by our experimental system corresponds to $\rho_0 \lambda_0^3 = 7.8$, with $\lambda_0 = 461~$nm the wavelength of the light used for imaging.

%


For a dense atomic cloud (i.e., not fulfilling the constraint given by Eq.~(\ref{eq:dense_regime})), Eq.~(\ref{eq:IndexRef}) was shown to be invalid by several experimental and theoretical studies \cite{Chomaz2012, Skipetrov2014, Guerin2017, Jennewein2018, Andreoli2021}. When two identical atoms are subject to incident quasiresonant light, being closer than one wavelength of the incoming radiation, they interact with it in an inherently collective way, due to an effective dipole-dipole interaction mediated by the vacuum modes of the electromagnetic radiation \cite{Lehmberg1970}. This effective interaction is included in the so-called coupled-dipole model\cite{Bienaime2011}, which describes the light propagation in the linear optics regime within an ensemble of classical electric dipoles. Several collective effects are captured by this model, such as subradiance\cite{Guerin2016, Cipris2021}, linear-optics superradiance\cite{Scully2016Directed,Araujo2016}, collective Lamb shift, reduction of the scattering coefficient \cite{Jennewein2018} and the index of refraction \cite{Andreoli2021} of the ensamble with respect to the mean-field result of Eq.~(\ref{eq:IndexRef}). In particular, the reduction of the scattering amplitude and the index of refraction of the atomic cloud are shown to scale with the dimensionless quantity $\rho_0 |\alpha (\Delta)|$. This implies that for Eq.~(\ref{eq:IndexRef}) to be valid, we must have $\rho_0 |\alpha (\Delta)|  \ll 1$, or equivalently
\begin{equation}
    \left|\Delta\right| \gg \frac{3}{8\pi^2} \,\rho_0 \,\lambda_0^3 \, \Gamma  \ .
    \label{eq:dense_condition}
\end{equation}
This condition ensures that the impact of collective effects on the refractive index is minimal even for spatially dense atomic samples. As a result, the refractive index can be approximated as:
\begin{equation}
    n(\bm{r},\Delta) \simeq 1 + \frac{\rho(\bm{r}) \alpha(\Delta)}{2}.
    \label{eq:n}
\end{equation}

In this regime, the index of refraction captures all the information to describe the propagation of  a laser field  through the cloud. The electric field of an incoming plane wave propagating in the $z$ direction is given by 
\begin{eqnarray}
    \nonumber \bm{E}(\bm{r},\Delta) &=& \bm{E}_0 \, \e^{i k \int_{-\infty}^z n(\bm{r},\Delta)\, \deriv z} \, \e^{ -i \omega t}  \\
    \label{eq:E} &=& \bm{E}_0  \, \e^{i k \int_{-\infty}^z \left[n(\bm{r},\Delta)-1\right] \, \deriv z} \, \e^{i k z -i \omega t} \\ 
    \nonumber &=& \e^{i k \int_{-\infty}^z \left[n(\bm{r},\Delta)-1\right] \, \deriv z} \, \bm{E}_i(\bm{r}) \ ,
\end{eqnarray}
 with $k = \omega/c$ the wavenumber of light in vacuum and $\bm{E}_0$ the complex amplitude of the electric field before passing through the atomic cloud (in the limit $z \rightarrow - \infty$). In the last term of the above equation, we have explicitly separated the effect of the index of refraction from the solution for light propagating in vacuum, $\bm{E}_i(\bm{r}) = \bm{E}_0 \, \e^{i k z -i \omega t}$, i.e. the incoming laser field.

The index of refraction presents an imaginary part $n_i$, related to the light absorption in the material, and a real part $n_r$, related to the extra phase accumulated by the light propagating within the material medium:
\begin{equation}
    n(\bm{r},\Delta) = n_r (\bm{r},\Delta) + i\, n_i (\bm{r},\Delta)  \ .
\end{equation}




If Eq.~(\ref{eq:dense_condition}) is valid, we can write
\begin{equation}
    \bm{E}_t(\bm{r},\Delta) = \e^{-\frac{b(x,y,\Delta)}{2}} \, \e^{i \phi(x,y,\Delta)} \, \bm{E}_i(\bm{r})
\end{equation}
with
\begin{eqnarray}
    \nonumber b(x,y,\Delta) &=& 2k \int_{-\infty}^{\infty} n_i (\bm{r},\Delta) \, \deriv z \\
    &=& \sigma_0 \frac{1}{1 + 4 \frac{\Delta^2}{\Gamma^2}}  \rho_{\text{2D}} (x,y)
    \label{eq:optical_depth}
\end{eqnarray}
the optical depth of the atomic sample, responsible for the attenuation of the intensity; and
\begin{eqnarray}
    \nonumber \phi(x,y,\Delta) &=& k \int_{-\infty}^{\infty} \left[n_r (\bm{r},\Delta) - 1 \right] \,\deriv z \\ 
    &=& \sigma_0 \frac{ - \frac{\Delta}{\Gamma}}{1 + 4 \frac{\Delta^2}{\Gamma^2}} \rho_{\text{2D}} (x,y)
    \label{eq:phase_shift}
\end{eqnarray}
the extra phase accumulated by the light due to the presence of the atoms. Both the optical depth and phase shift depend on the density of the sample integrated along the direction of the light propagation
\begin{equation}
    \rho_{\text{2D}} (x,y) = \int_{-\infty}^{\infty} \rho (x,y,z) \, \deriv z \ 
\end{equation}
and on the scattering cross-section of light at resonance
\begin{equation}
    \sigma_0 = \frac{3 \lambda_0^2}{2 \pi} \ .
    \label{eq:cross_section}
\end{equation}

Their product defines the so-called resonant optical depth:

\begin{equation}
    b_0(x,y)=b(x,y,\Delta=0)=\sigma_0 \rho_{\text{2D}} (x,y).
    \label{eq:resonant_optical_depth}
\end{equation}

\







For dense clouds, Eq.~(\ref{eq:optical_depth}) is only valid for detunings satisfying Eq.~(\ref{eq:dense_condition}), indicating that the resonant optical depth is no longer directly proportional to the 2D density. On the other hand, for $\Delta \ne 0$, the phase shift $\phi(x,y,\Delta)$ induced by the atomic distribution is different from zero, and the inhomogeneous transverse distribution of the phase shift converts the atomic density profile into a lens for the incident intensity. This lensing effect \cite{Bradley1995, Andrews1997} diffracts the light, particularly at the edges of the ensemble where the variation of $\rho_{\text{2D}}$ is steeper. This results in a false wave-like pattern in the measured 2D density profile, leading to an inaccurate measurement of the size of the atomic cloud.
This effect is generally unavoidable in absorption imaging with a detuned, broad laser beam for small atomic clouds. Consequently, absorption measurements of the optical depth far from resonance are typically used only to probe the local optical depth with a beam much smaller than the cloud's transverse size. This ensures that the phase shift remains uniform over the light’s wavefront, but sacrifices information on the full 2D atomic density profile.
We also note that, although absorption imaging with a saturated incident intensity \cite{Reinaudi2007} yielded promising results for dilute clouds in the high optical depth regime, the same limits in terms of density must be applied to this technique. 
 In summary, the above discussion demonstrates that it is not possible to measure $\rho_{\text{2D}}$ through direct optical depth measurements.

\subsection{PCI of the dense atomic cloud}

The PCI protocol aims at obtaining $\rho_{\text{2D}} (x,y)$ by measuring the spatially resolved phase shift induced by the atoms on the incoming light. In order to implement this protocol, the detuning $\Delta$ of the laser is selected to ensure minimal light absorption. This is necessary, since the phase shift will be cast onto an intensity modulation (as shown later in Eq.~(\ref{eq:PCI_extract})) that would be difficult to interpret in presence of absorption. For a dilute atomic sample with a high maximum resonant optical depth, $b_0 \equiv \text{max}(b_0(x,y)) \gg 1$, the condition for negligible light absorption $b(x,y,\Delta) \ll 1$ translates to 
\begin{equation}
    \left|\Delta\right| \gg \sqrt{b_0} \, \frac{\Gamma}{2} \ .
    \label{eq:abs_negligible}
\end{equation}

By comparing Eqs.~(\ref{eq:phase_shift}) and (\ref{eq:resonant_optical_depth}),  we can express the phase shift, in the limit of large detunings, as a function of the resonant optical depth of the sample as

\begin{equation}
    \phi(x,y,\Delta) \simeq - \frac{\Gamma}{4\Delta} \sigma_0 \rho_{\text{2D}} (x,y) =- \frac{\Gamma}{4\Delta} \, b_0(x,y)  \ .
    \label{eq:phase_density}
\end{equation}

To have a good signal-to-noise ratio in the measurement of the phase shift, its maximum value must be at least on the order of one, $\phi_{\text{max}} (\Delta) \sim 1$. For a cloud with $b_0 \gg 1$, this condition is restated as 
\begin{equation}
    \left|\Delta \right|\sim b_0 \, \frac{\Gamma}{4} \ .
    \label{eq:Delta_PCI}
\end{equation}
Note that for $b_0 \gg 1$, the condition (\ref{eq:Delta_PCI}) already implies condition (\ref{eq:abs_negligible}).

One could also ask how the condition (\ref{eq:Delta_PCI}) relates to Eq.~(\ref{eq:dense_condition}), that must be satisfied when imaging spatially dense clouds. 
For atomic samples commonly encountered in cold atom experiments, trapped in harmonic or uniform traps, we have $b_0 \sim \sigma_0 \rho_0 \, \Delta z$, with $\Delta z$ the typical size of the cloud in the $z$ direction. Using Eq.~(\ref{eq:cross_section}), we conclude that a detuning satisfying Eq.~(\ref{eq:Delta_PCI}) will also satisfy Eq.~(\ref{eq:dense_condition}) if
\begin{equation}
    \Delta z \gg \frac{1}{\pi} \lambda_0 \ ,
    \label{eq:PCI_size}
\end{equation}
a condition fulfilled by most atomic samples produced in the laboratory.
Both conditions (\ref{eq:Delta_PCI}) and (\ref{eq:PCI_size}) ensure the linear relation between the measured $\phi(x,y,\Delta)$ and the 2D density profile $\rho_{\text{2D}}(x,y)$ given by Eq.~(\ref{eq:phase_shift}), which is fundamental for reliably reconstructing the 2D density profile of the atomic cloud from the PCI imaging.

\subsection{Extracting the atomic density profile from the PCI}

In order to measure the phase shift induced by the cloud, one needs to perform an interferometric measurement of the phase imprinted on the light wavefront. In the next section, we will see that it is possible to experimentally separate the incoming electric field from the electric field diffracted by the atoms. Following this approach, we express the transmitted electric field in the PCI setup as a combination of the incoming and the scattered field, 
\begin{eqnarray}
     \nonumber \bm{E}_{t,\text{PCI}}(\bm{r},\Delta)& =&  \e^{i \phi(x,y,\Delta)} \, \bm{E}_i(\bm{r}) \\
     &=& \bm{E}_i(\bm{r}) + \left(\e^{i \phi(x,y,\Delta)} - 1 \right) \bm{E}_i(\bm{r}) \label{eq:PCI_extract}\\
     \nonumber &=&  \bm{E}_i(\bm{r}) + \bm{E}_d(\bm{r},\Delta)
\end{eqnarray}
with the scattered electric field
\begin{equation}
     \bm{E}_d(\bm{r},\Delta) = \left(\e^{i \phi} - 1 \right) \bm{E}_i(\bm{r}) 
\end{equation}
and with $\phi \equiv \phi(x,y,\Delta$). A common technique to obtain the phase shift involves applying a controlled global phase shift $\phi_P$ to the non-scattered laser field. As a result, the total laser field $\bm{E}_{c}$ becomes
\begin{eqnarray}
    \nonumber \bm{E}_{c} &=& \e^{i \phi_P} \bm{E}_i(\bm{r}) + \bm{E}_d(\bm{r},\Delta) \\
    &=& \left(\e^{i \phi_P} + \e^{i \phi} - 1 \right) \bm{E}_i(\bm{r}) \ .
    \label{eq:E_c}
\end{eqnarray}
Finally, the corresponding intensity, which can be measured experimentally with a camera, normalized by the incoming intensity, is equal to 
\begin{equation}
    \frac{I_c}{I_0} = \frac{\left|\bm{E}_{c}\right|^2}{\left|\bm{E}_{i}\right|^2} = 3 - 2 \cos \phi_P + 4 \sin \frac{\phi_P}{2} \sin \left(\phi -\frac{\phi_P}{2} \right)\ .
    \label{eq:I_c}
\end{equation}

\begin{figure}
\centering
\includegraphics[width =\columnwidth]{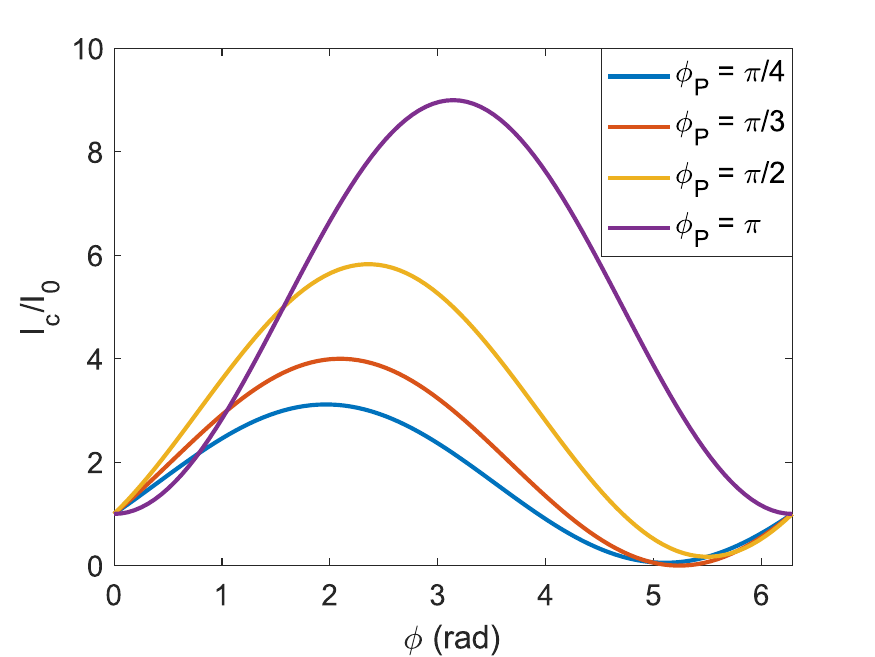}%
\caption{\label{fig:fig1} Intensity detected with the PCI imaging system, expressed in units of incident intensity, as a function of the phase shift $\phi$ induced by the atoms, for different controlled phase shifts $\phi_P$ applied to the incoming laser field component of the transmitted light.}%
\end{figure}

In Fig.~\ref{fig:fig1} we show the ratio $I_c/I_0$ as a function of $\phi$ for several values of $\phi_P$. The phase shift $\phi_P$ is chosen based on several criteria. On one hand, the phase $\phi_P = \pi/3$ allows setting the minimum value of $I_c$ to zero. On the other hand, the phase $\phi_P = \pi$ guarantees a maximum absolute amplitude of $8 I_0$ between the maximum and minimum values of $I_c$. However, its response is quadratic for small  $\phi$, making it unsuitable for measuring small  $\phi$ values as it introduces excessive noise to the density measurement, especially at the tails of the density distribution.
Another important feature is that the dependence of $I_c$ on $\phi$ is periodic, meaning there is an infinite set of $\phi$ values that result in the same value of $I_c/I_0$. In the region $(\phi_P-\pi)/2 \le \phi \le (\phi_P+\pi)/2$, Eq.~(\ref{eq:I_c}) gives
\begin{equation}
    \phi \equiv \phi_0 = \arcsin \left[\frac{\frac{I_c}{I_0} +  2 \cos \phi_P - 3}{4 \sin \frac{\phi_P}{2}}\right] + \frac{\phi_P}{2} \ .
    \label{eq:PCI_phase}
\end{equation}
On the other hand, without any assumption on $\phi$, its value can be any of the values in the family
\begin{equation}
    \phi_n = n \pi + \frac{\phi_P}{2} + \left(-1\right)^n \left(\phi_0 - \frac{\phi_P}{2}\right) \ ,
    \label{eq:phase_wrap}
\end{equation}
where $n$ is an integer ($n \in \mathbb{Z}$). 
Thus, when the measured phases span a wide range of values,  determining $\phi$ as a function of the measured 
$I_c$ generally requires an algorithm to unwrap the phase, resolving any ambiguity in assigning the correct value from the set of possible $\phi$ values.

\section{Characterization of the PCI technique for a dense atomic cloud}
\label{sec:PCI_results}

\subsection{The implementation of the PCI setup}

\begin{figure} [h!]
\centering
\includegraphics[width=\columnwidth]{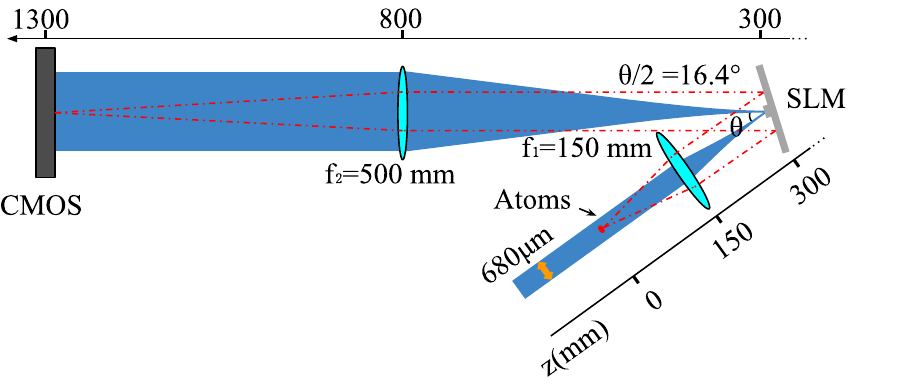}
\caption{\label{fig:PCI_setup} Experimental setup for the PCI implemented on our system. The blue shaded area represents the incoming imaging beam, while the red dot-dashed line represents the light scattered by the atoms and propagating to the camera.}%
\end{figure}

Following the theoretical discussion of Sec.~\ref{sec:PCI_theory}, we now describe the implementation of a versatile PCI setup for the characterization of our dense atomic cloud. A sketch of its optical setup is shown in Fig.~\ref{fig:PCI_setup}. The main difference compared to most implementations is the use of a reflective SLM (model Pluto2-VIS016 from Holoeye) at the focal plane of the light, instead of a passive phase plate. The SLM allows imprinting an arbitrary phase mask on the wavefront of the light, limited only by its resolution of $8~\upmu$m/pixel. In Fig.~\ref{fig:PCI_setup}, the center of the SLM is aligned with the center of the incoming beam, which is focused on the SLM by a lens of focal length $f_1 = 150~$mm. In this way, we can shift the phase of the nondiffracted light by applying a circular phase mask of radius $R_P$ localized at the center of the SLM. In the following, the phase difference that the phase mask imprints on the incoming light in comparison to the diffracted light is fixed to $\phi_P = \pi/3$ (see Eq.~(\ref{eq:E_c})).

\begin{figure} [h!]
\centering
\includegraphics[width=\columnwidth]{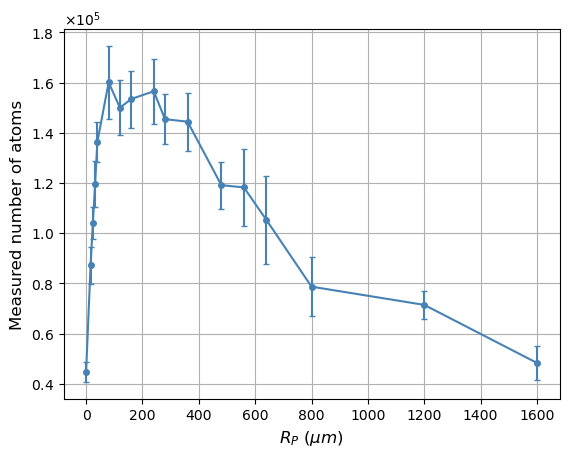}
\caption{\label{fig:R_scan} Number of atoms measured by imaging an atomic cloud at the end of the experimental procedure with the PCI setup, as a function of the radius of the phase mask disk.
}%
\end{figure}

The waist of the imaging beam ($\lambda_0 = 461$ nm) at the focal plane of the first lens, where the SLM is positioned, is related to the waist of the collimated beam at the atomic plane $W = 680~\upmu$m as $w_f = \lambda_0 f_1 /(\pi W) = 32~\upmu$m. To ensure that almost all non-diffracted laser intensity is confined within the disk, it is optimal to set the radius of the phase mask to $R_P \gtrsim 3 w_f = 100~\upmu$m. At the same time, in order to implement the interferometric measurement described by Eq.~(\ref{eq:E_c}), ideally no diffracted light should suffer a phase shift. The light diffracted by the atoms has a typical angular spread behind the atomic plane given by $\Delta \theta_a \sim 1/k R_r$; this angular spread is represented by the red dot-dashed line in Fig.~\ref{fig:PCI_setup}. The diffracted light is collimated by the first lens to a radius $f_1 \Delta \theta_a \sim f_1/(k R_r) = \lambda_0 f_1/(2 \pi R_r)$ at the SLM position; $R_P$ must thus also fulfill $R_P \ll \lambda_0 f_1/(2 \pi R_r) = 1.1~$mm. Both conditions are satisfied only if $W \gg R_r$, which is true in our setup. In this case, we expect a range of values of $R_P$ that fulfills both conditions. 

\begin{figure} [h!]
\centering
\includegraphics[width=\columnwidth]{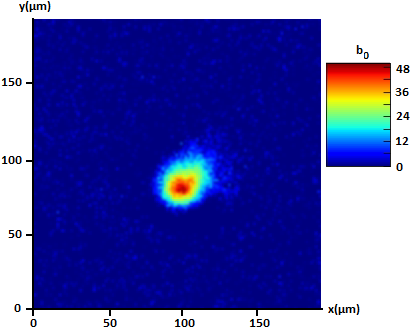}
\caption{\label{fig:image} 2D optical density profile of the \textit{in-situ} atomic cloud, directly measured by PCI with a probe light detuning of $-1010~$MHz, the phase shift $\phi_P=\pi/3$, and the radius of the phase mask $R_P=240~\upmu$m.}%
\end{figure}

In our setup, the flexibility of the SLM allows us to explore the impact of the parameter $R_P$ on the image of the atomic cloud captured by the CMOS camera. In Fig.~\ref{fig:R_scan}, we show the number of atoms measured \textit{in situ} for an atomic sample prepared by an experimental sequence similar to the one described in Sec.~\ref{sec:exp_seq}. To measure the number of atoms, we first convert the phase contrast imaging intensity profile into a phase profile following Eq.~(\ref{eq:PCI_phase}), then transform the latter into a 2D density profile via Eq.~(\ref{eq:phase_density}), as illustrated for example in Fig.~\ref{fig:image}. The 2D density profile is finally fitted by a 2D Gaussian profile, from which the total atom number is extracted. We identify a clear plateau for the number of atoms 
 within the range $100 \upmu\text{m} \lesssim R_P \lesssim 250 \upmu$m, compatible with the previously established condition for $R_P$. The number of atoms of the atomic cloud can also be directly compared to the number of atoms obtained from the absorption imaging after time of flight, which yielded $1.5 \times 10^5$ atoms — a value consistent with the one obtained at the plateau. For $R_P \lesssim 100~\upmu$m, the nondiffracted light is not fully confined within the phase mask, leading to a rapid decrease in the measured atom number as $R_P$ approaches zero. This behavior is expected, as without a phase mask, the light intensity at the image plane of the atomic position carries no information about the phase shifts induced by the atoms. On the other hand, for $R_P \gtrsim 250 \upmu$m, part of the diffracted light experiences a phase shift, causing the measured atom count to deviate from the anticipated value and eventually drop to zero as $R_P$ approaches infinity. This is also expected, since a global phase shift in the light results in the light intensity at the camera's position carrying no relevant information about the atomic cloud.

\begin{figure} [h!]
\centering
\includegraphics[width=\columnwidth]{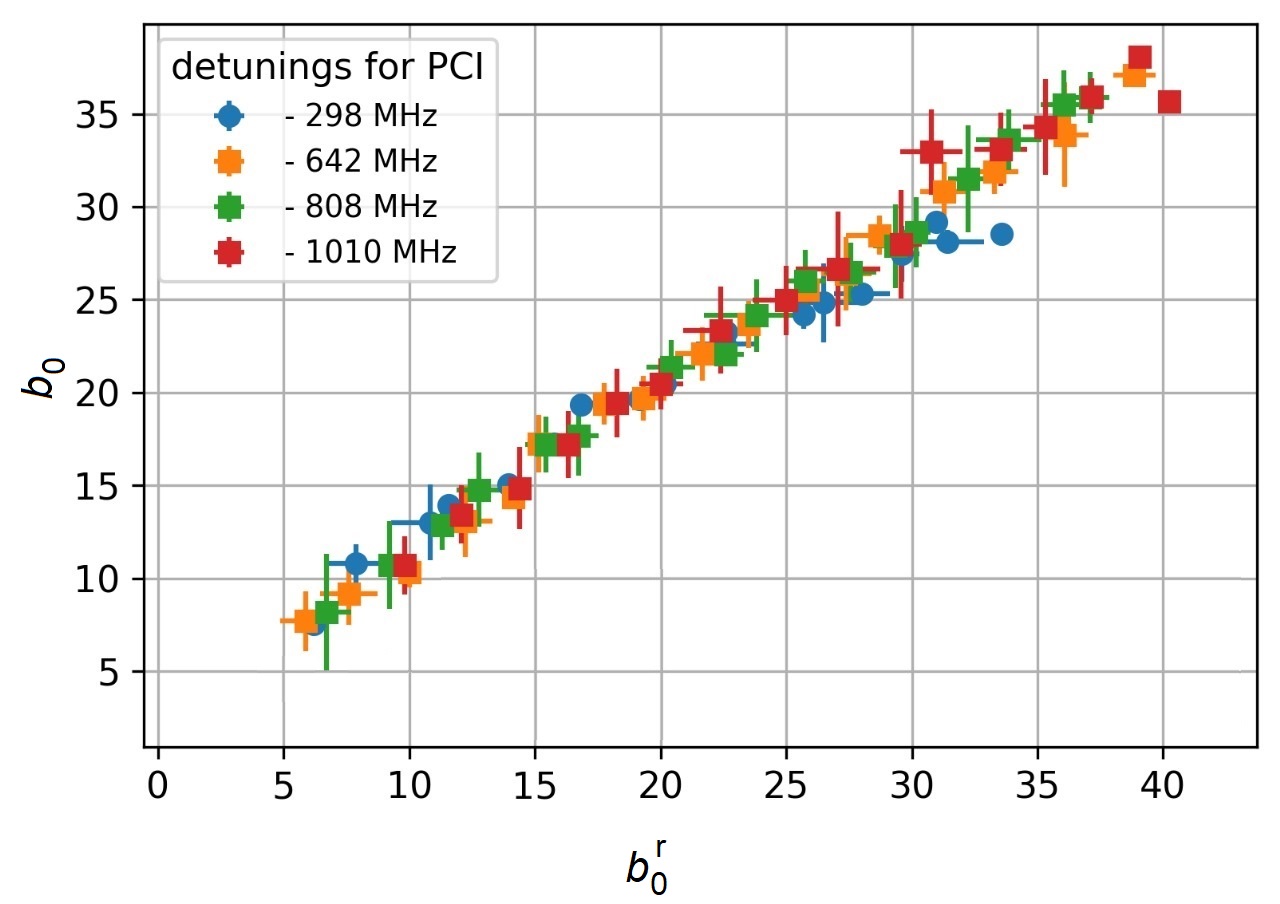}
\caption{\label{fig:OD} Resonant peak optical depth directly measured by PCI, as a function of the reconstructed peak optical depth calculated from the absorption imaging after some time of flight. Details in the main text.}%
\end{figure}

\subsection{PCI-reliable technique for in-situ density profiling of dense atomic clouds}

We now demonstrate that the PCI can be used to reliably determine the \textit{in-situ} density profile of a spatially and optically dense atomic cloud. To this end, we perform an \textit{in-situ} PCI, followed by an absorption imaging after a time of flight (ToF) of $\Delta t^{\text{ToF}} = 8~$ms on the same atomic cloud; this procedure relies on the fact that the first PCI image is non-destructive. This cloud was produced with the same experimental sequence described in Fig.~\ref{fig:experiment}(b) with a variable initial loading time of atoms in the blue MOT (i.e., shorter or equal to $1.5~$s). The variable loading time allows to load a different number of atoms in the final ODT, while maintaining a similar temperature for all loading times. From the \textit{in-situ} PCI we extract all relevant information about the transverse profile of the atomic cloud: $R_x$, $R_y$, $N$ and $b_0$, the resonant optical depth at the center of the atomic cloud (in the $z$ propagation direction of the imaging light). From the absorption imaging after time-of-flight, we can directly measure $N$ and the transverse sizes of the cloud after time-of-flight $R_x^{\text{ToF}}$ and $R_y^{\text{ToF}}$ in the $x$ and $y$ directions respectively. From the sizes after time of flight and supposing that the expanded size is much bigger than the \textit{in-situ} size, we obtain the temperature of the cloud,
\begin{equation}
T = \frac{m}{k_B} \frac{\left(R_x^{\text{ToF}}\right)^2 + \left(R_y^{\text{ToF}}\right)^2}{2 \left(\Delta t^{\text{ToF}}\right)^2},
\end{equation}
 where $m = 1.46 \cdot 10^{-25}~$kg is the mass of a single $^{88}$Sr atom. Using the temperature and the frequencies of the trap, we can reconstruct the \textit{in-situ} sizes of the Gaussian cloud as
\begin{equation}
R_{x,y}^{\text{r}} = \frac{1}{\omega_{x,y}} \sqrt{\frac{k_B T}{m}} \ .
\end{equation}
Combined with the measured number of atoms $N$, this allows us to determine the central in-situ optical depth
\begin{equation}
b_0^{\text{r}} = \sigma_0\frac{N}{2 \pi R_{x}^{\text{r}}  R_{y}^{\text{r}} } \ .
\end{equation}

This procedure based on the absorption imaging after time-of-flight, that relies on the trapping frequencies, is what is left to the experimentalist with no reliable \textit{in-situ} imaging available. Interatomic interactions, or out-of-equilibrium effects, that could induce deviations from the Gaussian density profile of a thermalized cloud, would not be detected from this indirect reconstruction method. On the other hand, any direct \textit{in-situ} imaging that is not diffraction-limited will provide reliable information on the exact transverse density profile of the cloud, in particular its optical depth.
In the particular case of the calibration discussed here, the deviations from a Gaussian profile are small, and we thus expect the two results to be in agreement. In Fig.~\ref{fig:OD} we plot $b_0$ directly measured by PCI for different values of the detuning $\Delta$ of the PCI light, as a function of $b_0^{\text{r}}$ reconstructed by absorption imaging on the same cloud. All the detunings respect the condition of Eq.~(\ref{eq:PCI_phase}), and are suited for performing the PCI. The slope of the curve determined by the points is close to 1 on the whole range, which shows that the PCI imaging faithfully reproduces the density profile of the cloud up to $b_0 = 40$, which corresponds to a maximum normalized density of $\rho_0 \lambda_0^3 = 4.9$, or $\eta = 0.37$. We note that for the smallest detuning of the PCI, $\Delta = -298~$MHz, the PCI measures a smaller $b_0$ than expected for the denser clouds. This is because our algorithm for interpreting the PCI does not unwrap the phase following Eq~(\ref{eq:phase_wrap}), considering we are always in the manifold with $n = 1$; this begins to be wrong for the right part of the graphic and $\Delta = -298~$MHz$~\sim -10 \Gamma$. Going to higher detunings (in absolute value) allows for faithfully measuring the density profile of clouds with higher $b_0$.
\\
\section{Conclusions}
\label{sec:conclusion} 

In this paper, we present a distinct approach for producing and characterizing a dense $^{88}$Sr atomic cloud, specifically designed to study light transport in dense atomic systems. A first major result of this work is the successful production of a cold $^{88}$Sr cloud with a resonant optical depth up to $b_0 = 64$, and spatial density $\rho_0 \lambda_0^3 = 7.8$. The key step in the procedure is the application of optical molasses on a narrow $^{88}$Sr transition, superimposed to the off-resonant optical dipole trap (ODT).
This molasses relies on a careful cancellation of any spurious magnetic field in the trapping region, and on the consideration of the effect of the AC Stark shift induced on the atomic transition by the ODT light. The narrow width of the transition is beneficial, in the sense that it allows us to partially circumvent the density limitations caused by the radiation pressure of quasi-resonant photons that are partially rescattered from the molasses. This is due to the important Doppler broadening of the transition, which, even at a temperature of $1~\upmu$K, is larger than the natural width of the transition.

We have then theoretically discussed the fundamental limitations for imaging the dense cloud, due to collective effects on the light absorption whenever $\rho_0 \lambda_0^3$ becomes of the order of 1. This discussion leads to two main conditions on the light detuning and on the longitudinal size of the atomic cloud, given respectively by Eqs. (\ref{eq:Delta_PCI}) and (\ref{eq:PCI_size}), that guarantee the validity of the PCI protocol we have implemented, despite the high spatial densities of our atomic sample. 
We also discuss our PCI setup, especially the use of SLM for imprinting the phase mask on the imaging light instead of a fixed phase plate, which allows the dynamic control and flexibility of the imaging system.

A central outcome of this work is the demonstration that PCI can reliably extract accurate 2D density profiles, even in high optical depth conditions, where traditional imaging techniques fail.
We show that the density profile of our atomic cloud extracted with this protocol is accurate for all ranges of $b_0$ and $\rho$ achievable within our experimental setup, allowing us to directly obtain the \textit{ in situ} 2D density profile of our atomic cloud.
This result is important because it establishes PCI as a viable and powerful tool for other experiments involving high-density samples. Researchers can now confidently apply PCI to characterize dense atomic systems. The dense atomic cloud, and the PCI imaging within the conditions developed here, will be fundamental for the main scientific goal of our experiment, which is to study light transport through a dense atomic cloud. In particular, in this rather unexplored regime, we expect to unravel interesting new physics on the crossover of many-body problems \cite{Andreoli2021} and the Anderson localization of light \cite{Skipetrov2016, Cottier2019}.

\section{Acknowledgements}
RCT acknowledges funding from the Brazilian funding agency FAPESP via grants 2018/23873-3 and 2013/07276-1; co-funding from FAPESP and French agency ANR via grant 2019/13143-0; funding from Brazilian CNPq projects 424865/2018-1, 406034/2021-4 and 313241/2023-6; agency PhWC acknowledges funding from the Brazilian funding agency FAPESP via grant 2013/04162-5; AC acknowledges funding from the Brazilian funding agency FAPESP via grant 2024/13462-7. M. H. acknowledges the French ANR (ANR19-CE47-0014-
01), the French government, through the UCAJ.E.D.I. Investments in the Future project managed by the National Research Agency (ANR) with the reference number ANR-15-IDEX-01, the STIC-AmSud (Ph879-17/CAPES
88887.521971/2020-00), from CAPES COFECUB
(Ph 997/23, CAPES 88887.711967/2022-00),
and the QUANTERA ERA-NET within the European Union’s Horizon 2020 Programme
(PACE-IN, 8C20004, ANR19-QUAN-003-01).

\section{Bibliography}

\bibliography{bibli}

\end{document}